\documentclass[prc,aps,twocolumn,showpacs]{revtex4}
\usepackage{amsmath,amssymb}
\usepackage{graphicx}
\usepackage{subfigure}
\def\pcomma{$^,$}
\def\pmainz{$^1$}
\def\pglsg{$^2$}
\def\pkent{$^3$}
\def\pbonn{$^4$}
\def\pgatch{$^5$}
\def\pgiess{$^6$}
\def\ppavia{$^7$}
\def\pgwu{$^8$}
\def\pucla{$^9$}
\def\plpi{$^{10}$}
\def\pdalh{$^{11}$}
\def\psaint{$^{12}$}
\def\ptomsk{$^{13}$}
\def\pedinb{$^{14}$}
\def\psackv{$^{15}$}
\def\plund{$^{16}$}
\def\pbasel{$^{17}$}
\def\pinr{$^{18}$}
\def\pzagreb{$^{19}$}
\def\pcua{$^{20}$}
\begin{document}

\title{
A new determination of the $\eta$ transition form factor in the Dalitz
decay $\eta \to e^+e^-\gamma$ with the Crystal Ball/TAPS detectors
at the Mainz Microtron}

\author{
P.~Aguar-Bartolom\'e\pmainz,
J.~R.~M.~Annand\pglsg,
H.~J.~Arends\pmainz,
K.~Bantawa\pkent,
R.~Beck\pbonn,
V.~Bekrenev\pgatch,
H.~Bergh\"auser\pgiess,
A.~Braghieri\ppavia,
W.~J.~Briscoe\pgwu,
J.~Brudvik\pucla,
S.~Cherepnya\plpi,
R.~F.~B.~Codling\pglsg,
C.~Collicott\pdalh\pcomma\psaint,
S.~Costanza\ppavia,
A.~Denig\pmainz,
E.~J.~Downie\pmainz\pcomma\pgwu,
P.~Drexler\pgiess,
L.~V.~Fil'kov\plpi,
A.~Fix\ptomsk,
D.~I.~Glazier\pedinb,
R.~Gregor\pgiess,
D.~J.~Hamilton\pglsg,
E.~Heid\pmainz\pcomma\pgwu,
D.~Hornidge\psackv,
L.~Isaksson\plund,
I.~Jaegle\pbasel,
O.~Jahn\pmainz,
T.~C.~Jude\pedinb,
V.~L.~Kashevarov\pmainz\pcomma\plpi,
I.~Keshelashvili\pbasel,
R.~Kondratiev\pinr,
M.~Korolija\pzagreb,
M.~Kotulla\pgiess,
A.~Koulbardis\pgatch,
S.~Kruglov\pgatch,
B.~Krusche\pbasel,
V.~Lisin\pinr,
K.~Livingston\pglsg,
I.~J.~D.~MacGregor\pglsg,
Y.~Maghrbi\pbasel,
D.~M.~Manley\pkent,
P.~Masjuan\pmainz,
J.~C.~McGeorge\pglsg,
E.~F.~McNicoll\pglsg,
D.~Mekterovic\pzagreb,
V.~Metag\pgiess,
A.~Mushkarenkov\ppavia,
B.~M.~K.~Nefkens\pucla,
A.~Nikolaev\pbonn,
R.~Novotny\pgiess,
H.~Ortega\pmainz,
M.~Ostrick\pmainz,
P.~Ott\pmainz,
P.~B.~Otte\pmainz,
B.~Oussena\pmainz\pcomma\pgwu,
P.~Pedroni\ppavia,
F.~Pheron\pbasel,
A.~Polonski\pinr,
S.~Prakhov\pmainz\pcomma\pgwu\pcomma\pucla\footnote[1]{corresponding author; e-mail: prakhov@ucla.edu},
J.~Robinson\pglsg,
G.~Rosner\pglsg,
T.~Rostomyan\pbasel,
S.~Schumann\pmainz,
M.~H.~Sikora\pedinb,
D.~I.~Sober\pcua,
A.~Starostin\pucla,
I.~I.~Strakovsky\pgwu,
I.~M.~Suarez\pucla,
I.~Supek\pzagreb,
C.~M.~Tarbert\pedinb,
M.~Thiel\pgiess,
A.~Thomas\pmainz,
M.~Unverzagt\pmainz\pcomma\pbonn\footnote[2]{corresponding author; e-mail: unvemarc@kph.uni-mainz.de},
D.~P.~Watts\pedinb,
D.~Werthm\"uller\pbasel,
and F.~Zehr\pbasel
\\
\vspace*{0.1in}
(A2 Collaboration at MAMI)
\vspace*{0.1in}
}

\affiliation{
\pmainz Institut f\"ur Kernphysik, University of Mainz,
D-55099 Mainz, Germany}
\affiliation{
\pglsg Department of Physics and Astronomy, University of Glasgow, Glasgow G12 8QQ, 
United Kingdom}
\affiliation{
\pkent Kent State University, Kent, Ohio 44242-0001, USA}
\affiliation{
\pbonn Helmholtz-Institut f\"ur Strahlen- und Kernphysik, University of Bonn, 
D-53115 Bonn, Germany}
\affiliation{
\pgatch Petersburg Nuclear Physics Institute, 188350 Gatchina, Russia}
\affiliation{
\pgiess II Physikalisches Institut, University of Giessen, D-35392 Giessen, Germany}
\affiliation{
\ppavia INFN Sesione di Pavia, I-27100 Pavia, Italy}
\affiliation{
\pgwu The George Washington University, Washington, DC 20052-0001, USA}
\affiliation{
\pucla University of California Los Angeles, Los Angeles, California 90095-1547, USA}
\affiliation{
\plpi Lebedev Physical Institute, 119991 Moscow, Russia}
\affiliation{
\pdalh Dalhousie University, Halifax, Nova Scotia B3H 4R2, Canada}
\affiliation{
\psaint Saint Mary's University, Halifax, Nova Scotia B3H 3C3, Canada}
\affiliation{
\ptomsk Laboratory of Mathematical Physics, Tomsk Polytechnic University, 634050 Tomsk, Russia}
\affiliation{
\pedinb School of Physics, University of Edinburgh, Edinburgh EH9 3JZ, United Kingdom}
\affiliation{
\psackv Mount Allison University, Sackville, New Brunswick E4L 1E6, Canada}
\affiliation{
\plund Lund University, SE-22100 Lund, Sweden}
\affiliation{
\pbasel Institut f\"ur Physik, University of Basel, CH-4056 Basel, Switzerland}
\affiliation{
\pinr Institute for Nuclear Research, 125047 Moscow, Russia}
\affiliation{
\pzagreb Rudjer Boskovic Institute, HR-10000 Zagreb, Croatia}
\affiliation{
\pcua The Catholic University of America, Washington, DC 20064, USA}

\date{\today}
         
\begin{abstract}
 The Dalitz decay $\eta \to e^+e^-\gamma$ has been measured
 in the $\gamma p\to \eta p$ reaction
 with the Crystal Ball and TAPS multiphoton spectrometers,
 together with the photon tagging facility at the Mainz Microtron MAMI.
 The experimental statistic used in this work is
 one order of magnitude greater than in any previous measurement
 of $\eta \to e^+e^-\gamma$.
 The value obtained for the slope parameter $\Lambda^{-2}$
 of the $\eta$ transition form factor,
 $\Lambda^{-2}=(1.95\pm 0.15_{\mathrm{stat}}\pm 0.10_{\mathrm{syst}})$~GeV$^{-2}$,
 is in good agreement with recent measurements conducted
 in $\eta \to e^+e^-\gamma$ and $\eta \to \mu^+\mu^-\gamma$ decays,
 as well as with recent form-factor calculations.
 The uncertainty obtained in the value of $\Lambda^{-2}$
 is lower compared to results from previous measurements
 of the $\eta \to e^+e^-\gamma$ decay.
\end{abstract}


\maketitle

\section{Introduction}

 The determination of electromagnetic transition form factors (TFFs)
 of light pseudoscalar mesons $P$ is of crucial importance for understanding
 the intrinsic structure of these particles (see Ref.~\cite{TFFW12} and references
 therein). A special role is occupied by the $P \to \gamma^* \gamma \to e^+ e^- \gamma$
 decays of light neutral mesons~\cite{Landsberg}. As this decay includes only one
 hadron, the TFF fully describes the electromagnetic structure of the particle.
 For structureless mesons, the decay rate can be calculated within
 Quantum Electrodynamics (QED). The complex internal structure of mesons
 incorporated in the TFF modifies the decay rate. By measuring the decay rate
 and dividing it by the QED contribution, the TFF can be determined.
\begin{figure}
\centering
\subfigure[]{
\includegraphics[height=3.2cm,width=3.8cm]{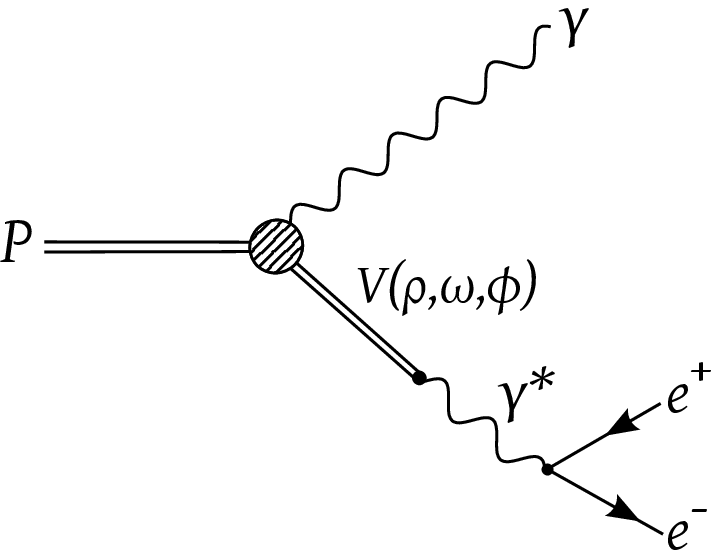}
\label{fig:feynman1:a}
}
\subfigure[]{
\includegraphics[height=3.2cm,width=3.8cm]{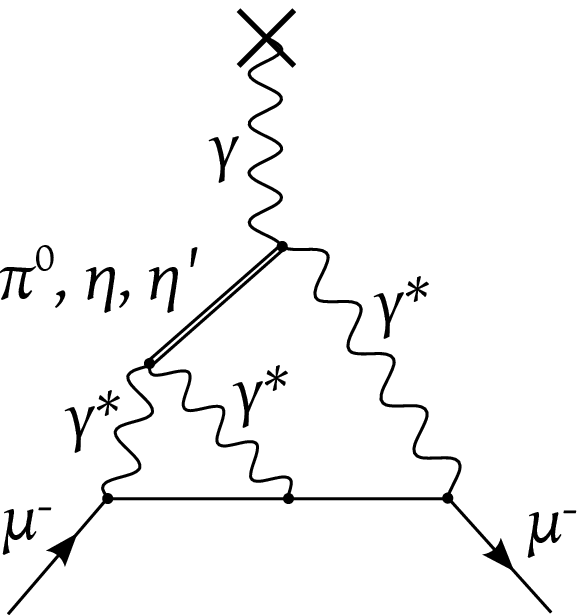}
\label{fig:feynman1:b}
}
\caption{
 Feynman diagrams illustrating (a) the $P \to \gamma^* \gamma$ transition in VMD
 and (b) one of the leading contributions to
 the correction for $(g-2)_\mu$ due to the hadronic light-by-light scattering.
}
\label{fig:feynman1}
\end{figure}
\begin{figure*}
\centering
\subfigure[]{
\includegraphics[height=3.4cm,width=5.cm]{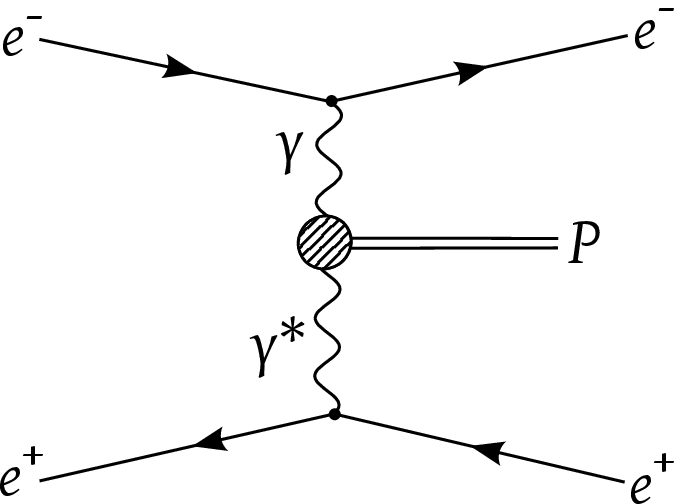}
\label{fig:feynman2:a}
}
\subfigure[]{
\includegraphics[height=3.4cm,width=4.5cm]{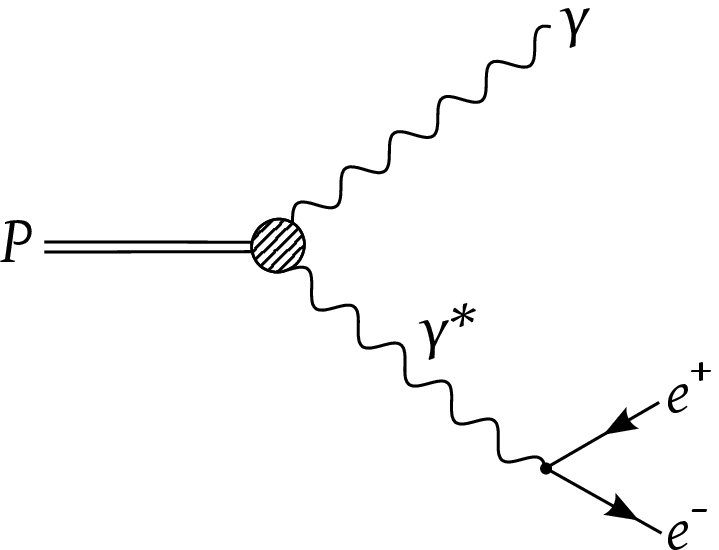}
\label{fig:feynman2:b}
}
\subfigure[]{
\includegraphics[height=3.4cm,width=5.5cm]{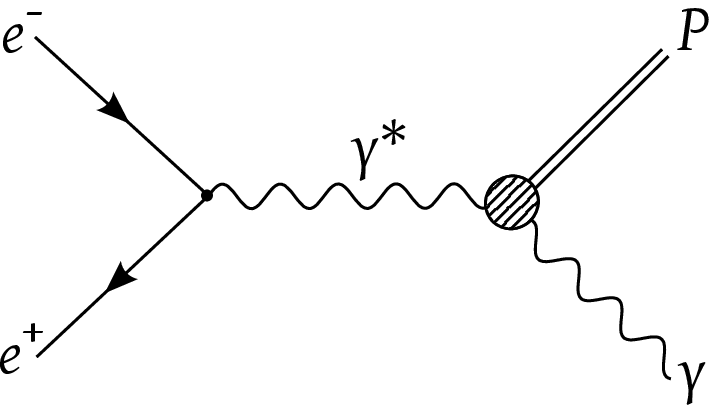}
\label{fig:feynman2:c}
}
\caption{
 Feynman diagrams showing the processes with the $P \to \gamma^* \gamma$
 transition that can be used for investigating TFFs
 of light pseudoscalar mesons $P$ in three different $q^2$ regions:
 (a)~spacelike ($q^2 <0$~GeV$^2$), (b)~timelike ($(2m_l)^2 < q^2 < m_P^2$),
 and (c)~timelike ($q^2 > m_P^2$).
}
\label{fig:feynman2}
\end{figure*}

 Decays of light pseudoscalar mesons ($\pi^0$, $\eta$, and $\eta'$)
 into a real and a virtual photon are ideally suited for testing the
 Vector Meson Dominance (VMD) model~\cite{Sakurai}.
 In VMD, the coupling of a virtual photon to a pseudoscalar meson is described via
 an intermediate virtual vector meson $V$ (see Fig.~\ref{fig:feynman1:a}).
 This mechanism is especially strong in the timelike momentum-transfer region,
 where a resonancelike behavior near momentum transfer $q^2 = m^2_V$ of the virtual photon
 arises because the virtual vector meson reaches the mass shell~\cite{Landsberg}.

 The TFFs of the $\eta$ and $\eta'$ mesons are strongly related to the mixing
 of these particles~\cite{Esc05}. With quarks as inner degrees of freedom in
 Quantum Chromodynamics (QCD), several symmetry-breaking mechanisms (U(1) axial anomaly
 and dynamical and explicit chiral-symmetry breaking) lead to a mixing of the pure SU(3)
 states $|\eta_q> = \frac{1}{\sqrt{2}} | u\bar u + d \bar d>$ and
 $|\eta_s> = | s \bar s>$ in the quark-flavor basis to form the $\eta$ and $\eta'$ mesons.
 In the picture of VMD, the coupling of light vector-meson resonances ($\omega$,
 $\rho$, and $\phi$) to the virtual photon influences the slope of the TFF.
 Since the $\phi$ meson is a pure $s \bar s$ state, measuring the TFF precisely gives
 stringent constraints on the strange-quark content of the $\eta$ and
 $\eta'$ mesons~\cite{Landsberg}.
 In addition, a possible gluonic contribution to $\eta$ and $\eta'$ is currently
 under discussion~\cite{Esc07}, which could be investigated with TFFs.
 If such an effect existed, it would have to be included in $\eta$-$\eta'$ mixing
 schemes. Exploring the TFFs of the $\eta$
 and $\eta'$ mesons sheds new light not only on the quark structure of these
 particles but also on the interplay of the symmetry-breaking mechanisms
 that were mentioned above.

 It has been argued~\cite{g_2} that TFFs might be related to the contribution of
 the hadronic light-by-light scattering 
 to the anomalous magnetic moment of the muon, $(g-2)_\mu$, which is one
 of the limiting contributions for the Standard Model (SM) calculation of
 this precision observable. Since the contribution of the hadronic light-by-light
 scattering can not be accessed directly through experiments,
 models describing $P \gamma^* \gamma$ transitions
 are needed to calculate this contribution. A diagram showing one of the leading
 contributions to the correction due to the hadronic light-by-light scattering
 is depicted in Fig.~\ref{fig:feynman1:b}.
 Here two vertices appear where pseudoscalar mesons couple to virtual photons
 and the external magnetic field.
 Experiments can measure such vertices in other processes,
 e.g., decays of pseudoscalar mesons, and this way act as a testing ground for
 VMD or VMD-based models describing $P \gamma^* \gamma$ transitions
 (see Ref.~\cite{g_2} and references therein) and provide appropriate input
 parameters for these models. Thus, the models describing
 the $P \gamma^* \gamma$ transitions have to be tested as precisely
 as possible to reduce the uncertainty in the SM prediction for $(g-2)_\mu$.
 Calculations of the contribution from the hadronic light-by-light scattering
 to $(g-2)_\mu$ are performed
 in the spacelike momentum-transfer region, although it is also possible to do
 in the timelike regime.
  The dominant energy region for such calculations turns out to be the low-energy
 region up to 1~GeV$^2$, exactly where the slope of the TFF is the relevant parameter.
 Since the TFF is continuous at $q^2=0$~GeV$^2$, this value can be determined
 in both the momentum regions. It is currently not possible
 to reach the lowest $q^2$ values experimentally in the spacelike regime.
 On the contrary, determining TFFs with high precision is possible
 from Dalitz decays. Thus, measurements of the timelike TFF with Dalitz decays
 are important for fixing the slope at $q^2=0$~GeV$^2$, especially by
 studying decays like $\eta \to l^+l^-\gamma$ with the $e^+e^-$ lepton pair,
 making it possible to reach a much lower value of $q^2 = 4m_l^2$
 than with $\mu^+\mu^-$.

 Experimentally, TFFs can be explored through different techniques
 in three separate momentum-transfer regions. The spacelike region,
 $q^2 <0$~GeV$^2$, can be studied at an $e^+e^-$ collider
 via the $\gamma^* \gamma \to P$ process in the $e^+e^- \to e^+e^- P$ reaction
 (see Fig.~\ref{fig:feynman2:a}). The lowest timelike $q^2$ region,
 $(2m_l)^2 < q^2 < m^2_P$, is accessible only through studying meson decays,
 where $2m_l$ is the mass of the two leptons from the virtual-photon decay
 (see Fig.~\ref{fig:feynman2:b}). Above $q^2 = m^2_P$, TFFs can be investigated
 in $e^+e^-$-annihilation processes, $e^+e^- \to P \gamma$ (Fig.~\ref{fig:feynman2:c}),
 at collider experiments.

 In this work, a new determination of the timelike $\eta$ TFF
 in the low-$q^2$ region is presented. Experimentally, such a determination 
 can be done by measuring the decay rate
 of $\eta\to \gamma^*\gamma \to l^+l^-\gamma$ as a function of a dilepton
 invariant mass $m_{ll}=q$ and normalizing it to the partial decay
 width $\Gamma(\eta\to \gamma\gamma)$~\cite{Landsberg}:
\begin{eqnarray}
 & & \frac{d\Gamma(\eta\to l^+l^-\gamma)}{dm_{ll} \Gamma(\eta\to \gamma\gamma)} =
\nonumber 
\\ 
 & = & \frac{4\alpha}{3\pi m_{ll}} [1-\frac{4m^2_l}{m^2_{ll}}]^{\frac{1}{2}}
 \cdot [1+\frac{2m^2_l}{m^2_{ll}}] \cdot [1-\frac{m^2_{ll}}{m^2_{\eta}}]^{3}
 \cdot |F_{\eta}(m_{ll})|^2
\nonumber 
\\ 
 & = & [QED] \cdot |F_{\eta}(m_{ll})|^2,
\label{eqn:dgdm}
\end{eqnarray}
 where $F_{\eta}$ is the TFF of the $\eta$ meson and $m_{\eta}$ is the mass
 of the $\eta$ meson.
 Assuming VMD, transition form factors are usually parametrized as 
\begin{equation}
 F(m_{ll}) = \frac{1}{1-\frac{m^2_{ll}}{\Lambda^2}},
\label{eqn:Fm}
\end{equation}
 where $\Lambda$ is the effective mass of the virtual vector mesons.
 The parameter $b=\Lambda^{-2}$ reflects the form-factor slope at $m_{ll}=0$.
 A simple VMD model would incorporate only the $\rho$, $\omega$, and $\phi$
 resonances (in the narrow-width approximation) as virtual vector mesons
 driving a photon interaction with a pseudoscalar. Using a quark model to account
 for the corresponding couplings would yield the TFF slope
 $b = 1.8$~GeV$^{-2}$~\cite{Landsberg}, corresponding to $\Lambda = 745$~MeV.

 So far, the most precise measurement of the parameter $\Lambda^{-2}$
 in the Dalitz decay $\eta \to e^+e^-\gamma$ was reported in 2011
 by the A2 Collaboration at MAMI~\cite{CB_2011}.
 The value obtained,
 $\Lambda^{-2}=(1.92\pm 0.35_{\mathrm{stat}}\pm 0.13_{\mathrm{syst}})$~GeV$^{-2}$,
 is based on an analysis of $1.35\cdot 10^3$ $\eta \to e^+e^-\gamma$ decays. 
 At the same time, an analysis of the $\eta \to \mu^+\mu^-\gamma$ decay
 by the NA60 Collaboration allowed a determination of $\Lambda^{-2}$
 with significantly better statistical accuracy.
 In 2009, the NA60 Collaboration reported the value
 $\Lambda^{-2}=(1.95\pm 0.17_{\mathrm{stat}}\pm 0.05_{\mathrm{syst}})$~GeV$^{-2}$
 based on an analysis of $9\cdot 10^3$ $\eta \to \mu^+\mu^-\gamma$
 decays in peripheral In--In data~\cite{NA60_2009}.
 Recently, NA60 presented an improved preliminary result,
 $\Lambda^{-2}=(1.951\pm 0.059_{\mathrm{stat}}\pm 0.042_{\mathrm{syst}})$~GeV$^{-2}$,
 based on an analysis of $8\cdot 10^4$ low-mass muon
 pairs produced in $p-A$ collisions~\cite{NA60_2012}.
 
 The major shortcoming of using the $\eta \to \mu^+\mu^-\gamma$ Dalitz decay
 for the determination of the $\eta$ TFF is the inability
 to measure $d\Gamma(\eta\to l^+l^-\gamma)/dm_{ll}$ below $m_{ll}=2 m_{\mu}$.
 This limitation does not allow a check of whether the experimental
 points approach $F_{\eta}=1$ at $m_{ll}=0$, as it was
 assumed in their final fit. Besides, the measurement of the NA60 Collaboration
 is based on fitting all possible contributions to the entire spectrum of
 the $\mu^+\mu^-$ invariant masses, without detecting the final-state photon
 and reconstructing the $\eta$ meson from its decay products.

 Using the $\eta \to e^+e^-\gamma$ Dalitz decay for the determination of
 the $\eta$ TFF allows for measuring
 $d\Gamma(\eta\to l^+l^-\gamma)/dm_{ll}$ much closer to $m_{ll}=0$
 and to fit a function with two free parameters to the data points.
 One of the function parameters is $\Lambda^{-2}$ itself;
 the other reflects the uncertainty
 in the general normalization of the data points.
 Such uncertainty could emerge, for example,
 from the experimental determination of the number of
 $\eta$ mesons produced, which is needed
 for calculating $d\Gamma(\eta\to l^+l^-\gamma)/dm_{ll}$.

 The results of the work presented in this paper
 are based on an analysis of $2.2\cdot 10^4$
 $\eta \to e^+e^-\gamma$ decays from a total of $3 \cdot 10^7$ $\eta$
 mesons produced in the $\gamma p\to \eta p$ reaction.
 About one-third of these data have already been used in the previous
 analysis by the A2 Collaboration at MAMI~\cite{CB_2011}. 
 Compared to the analysis of Ref.~\cite{CB_2011}, further increase
 in statistic was also achieved through exploiting the full production
 energy range for $\eta$ mesons, using, unlike the former analysis,
 a kinematic-fit technique for event identification,
 and substantially revising the criteria used for event selection.

\section{Experimental setup}
\label{sec:Setup}

The process $\gamma p\to \eta p \to e^+e^-\gamma p$
was measured using the Crystal Ball (CB)~\cite{CB}
as the central spectrometer and TAPS~\cite{TAPS,TAPS2}
as a forward spectrometer. These detectors were
installed in the energy-tagged bremsstrahlung photon beam of
the Mainz Microtron (MAMI)~\cite{MAMI,MAMIC}. 
The photon energies were determined
by the Glasgow-Mainz tagging spectrometer~\cite{TAGGER2,TAGGER,TAGGER1}.

The CB detector is a sphere consisting of 672
optically isolated NaI(Tl) crystals, shaped as
truncated triangular pyramids, which point toward
the center of the sphere. The crystals are arranged in two
hemispheres that cover 93\% of $4\pi$ sr, sitting
outside a central spherical cavity with a radius of
25~cm, which is designed to hold the target and inner
detectors. In this experiment, TAPS was
arranged in a plane consisting of 384 BaF$_2$
counters of hexagonal cross section.
It was installed 1.5~m downstream of the CB center
and covered the full azimuthal range for polar angles
from $1^\circ$ to $20^\circ$.

The present analysis is based on the same data set that was
used to study the $\eta \to 3\pi^0$ decay~\cite{slopemamic}
and to measure the $\gamma p\to \eta p$ differential cross
sections~\cite{etamamic}.
The experiment was conducted in 2007 by using the 1508-MeV
electron beam from the Mainz Microtron, MAMI-C~\cite{MAMIC}.
Bremsstrahlung photons, produced by the 1508-MeV electrons
in a 10-$\mu$m Cu radiator and collimated by a 4-mm-diameter Pb collimator,
 were incident on a 5-cm-long liquid hydrogen (LH$_2$) target located
in the center of the CB.
The total amount of the material around the target, including the Kapton cell and
the 1-mm-thick carbonfiber beamline, was equivalent to  0.8\% of a radiation length $X_0$.
In the present measurement, it was essential to keep the material budget
 as low as possible to diminish the $\eta \to \gamma\gamma$ background with
 conversion of real photons into $e^+e^-$ pairs.

The energies of the incident photons were analyzed up to 1402~MeV by detecting
the postbremsstrahlung electrons in the Glasgow-Mainz tagger~\cite{TAGGER2}.
The energy resolution of the tagged photons is mostly defined by the width
of the tagger focal-plane detectors and by the electron-beam energy.
For the present beam energies, a typical width of a tagger channel was about 4~MeV.
\begin{figure}
\includegraphics[width=7.cm,height=8.cm,bbllx=0.cm,bblly=0.cm,bburx=10.5cm,bbury=12.cm]{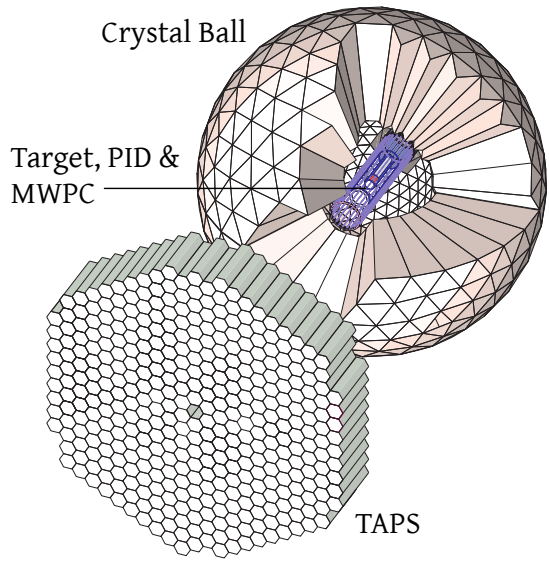}
\caption{(Color online)
  A general sketch of the CB, TAPS, and PID detectors.
}
 \label{fig:cb_taps_pid} 
\end{figure}

 The target was surrounded by a Particle IDentification
 (PID) detector~\cite{PID} used to distinguish between charged and
 neutral particles. It was made of 24 scintillator bars
 (50 cm long, 4 mm thick) arranged as a cylinder with a radius of 12 cm.

The experimental data were taken with a trigger that
required, first, the total-energy deposit in the CB
to exceed $\sim 320$~MeV and, second, the number of so-called hardware clusters
in the CB (multiplicity trigger) to be larger than two.
Depending on the data-taking period, events with cluster multiplicity two 
were prescaled with a different rate.
TAPS was not in the multiplicity trigger for these experiments.

 More details on the experimental resolutions of the detectors and
 other conditions during these measurements
 are given in Refs.~\cite{slopemamic,etamamic}. A general sketch
 of the CB, TAPS, and PID is shown in Fig.~\ref{fig:cb_taps_pid}.

\section{Data handling}
\label{sec:Data}

 Candidates for the process $\gamma p\to \eta p \to e^+e^-\gamma p$ were
 extracted from the analysis of events having
 three and four clusters reconstructed in the CB and TAPS together.
 The three-cluster events were analyzed assuming that the final-state
 proton was not detected.
 The selection of event candidates was based on
 the kinematic-fit technique.
 The details of the kinematic-fit parametrization
 of the detector information and resolution are given
 in Ref.~\cite{slopemamic}.
 Since electromagnetic (e/m) showers from electrons and positrons are
 very similar to those of photons, 
 the hypothesis $\gamma p \to 3\gamma p$ was tested to identify
 the $\gamma p\to e^+e^-\gamma p$ candidates.
 The events that satisfied this hypothesis with a probability greater than 2\%
 were accepted as possible reaction candidates.
 The kinematic-fit output was used to reconstruct the reaction kinematics.
 In this output, there was no identification of which e/m shower belonged
 to the outgoing photon, electron, or positron.  
 Since the main purpose of the experiment was to measure
 the $\eta \to e^+e^-\gamma$ decay rate as a function of
 the invariant mass $m(e^+e^-)$, the next step of the analysis was
 to separate the final-state photon from the electron and
 positron. This procedure was optimized by using a Monte
 Carlo (MC) simulation of the process $\gamma p\to \eta p \to e^+e^-\gamma p$.

 To reproduce the experimental yield of the $\eta \to e^+e^-\gamma$ decays
 depending on the incident-photon energy, the $\gamma p\to \eta p$ reaction
 was generated according to its excitation function,
 measured in the same experiment~\cite{etamamic},
 which was then folded with the bremsstrahlung energy dependence of the incident photons. 
 Since the energy range used in the analysis covers almost 700~MeV of the photon
 beam energies, the production angular distribution of $\gamma p\to \eta p$
 changes with energy. As this distribution averaged over all energies
 is sufficiently close to an isotropic distribution, for simplicity,
 the production angle was generated isotropically. 
 The $\eta \to e^+e^-\gamma$ decay was generated according to Eq.~(\ref{eqn:dgdm}),
 assuming the $\eta$ transition form factor $F_{\eta}=1$.

 Possible background processes were studied via MC simulation.
 The reaction $\gamma p\to \eta p$ was simulated for several other decay modes
 of the $\eta$ meson to check if they could mimic a peak
 from the $\eta \to e^+e^-\gamma$ signal.
 Such MC simulations were made for the $\eta\to\gamma\gamma$, $\eta\to\pi^0\pi^0\pi^0$,
 $\eta\to\pi^+\pi^-\pi^0$, and  $\eta\to\pi^+\pi^-\gamma$ decays.
 The energy dependence and the production angular distribution of
 all $\gamma p\to \eta p$ simulations were generated in the same way as 
 for the process $\gamma p\to \eta p \to e^+e^-\gamma p$. 
 In contrast to the $\eta \to e^+e^-\gamma$ decay, all other decays of $\eta$
 were generated according to phase space.
\begin{figure*}
\includegraphics[width=16.cm,height=5.75cm,bbllx=1.cm,bblly=.25cm,bburx=19.5cm,bbury=6.75cm]{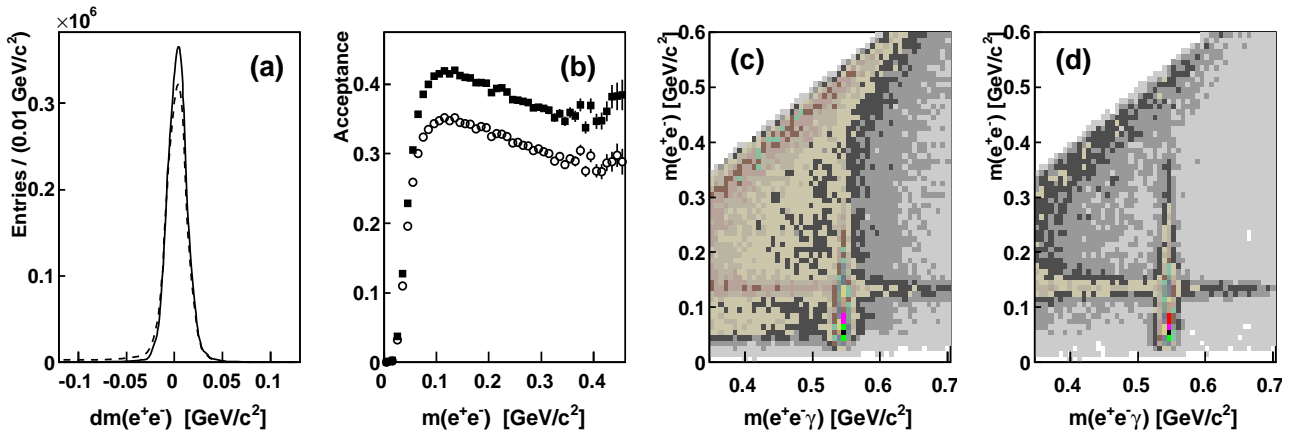}
\caption{ (Color online)~
 (a)~Difference between the generated and reconstructed
 invariant mass $m(e^+e^-)$ for the MC simulation of
 $\gamma p\to \eta p\to e^+e^-\gamma p$ shown for three- (dashed line) and four-cluster
 (solid line) events, where both the distributions are normalized to each other;
 (b)~The $\gamma p\to \eta p\to e^+e^-\gamma p$ acceptance as a function
 of $m(e^+e^-)$ shown for all (solid squares) and for only four-cluster (open circles)
 events;
 (c) and (d) Distributions of the invariant mass $m(e^+e^-)$ as a function of
 the invariant mass $m(e^+e^-\gamma)$ for the experimental
 $\gamma p\to e^+e^-\gamma p$ candidates shown for all and
 only four-cluster events, respectively. 
}
 \label{fig:eeg_fig1} 
\end{figure*}

 The major background under the peak from $\eta \to e^+e^-\gamma$ decays
 was found to be from the reaction $\gamma p\to \pi^0\pi^0p$.
 The MC simulation of this reaction was done in the same way as 
 reported in Ref.~\cite{p2pi0mamic}.
 Although this background is smooth in the region of the $\eta$ mass and 
 cannot mimic an $\eta \to e^+e^-\gamma$ peak, its MC simulation was
 used for optimizing the signal-to-background ratio and
 parametrizing the background under the signal.

 For all reactions, the simulated events
 were propagated through a {\sc GEANT} (version 3.21) simulation of the experimental
 setup. To reproduce resolutions of the experimental data,
 the {\sc GEANT} output (energy and timing) was subject
 to additional smearing, thus allowing both the simulated and experimental data
 to be analyzed in the same way.
 The simulated events were also tested to check whether they passed
 the trigger requirements.

 The optimization of other selection cuts was based on
 the analysis of the $\eta \to e^+e^-\gamma$ candidates
 selected with the kinematic fit to the experimental data and MC simulations.
 As it turned out,
 the $\gamma p\to e^+e^-\gamma p$ decays in the three-cluster sample
 (i.e., without the outgoing proton detected) have a level of
 background under the peak from $\eta \to e^+e^-\gamma$ that is significantly larger
 than in the four-cluster sample.
 This background was partially suppressed by testing
 the kinematic-fit hypotheses $\gamma p\to \pi^0p\to  \gamma\gamma p$
 and $\gamma p\to \eta p\to  \gamma\gamma p$ to the same events,
 then rejecting those for which the probability to be $\gamma p\to \pi^0p$
 or $\gamma p\to \eta p$ was greater than $10^{-6}$ and $10^{-5}$, respectively.
\begin{figure*}
\includegraphics[width=15.5cm,height=5.5cm,bbllx=1.cm,bblly=.25cm,bburx=19.5cm,bbury=6.75cm]{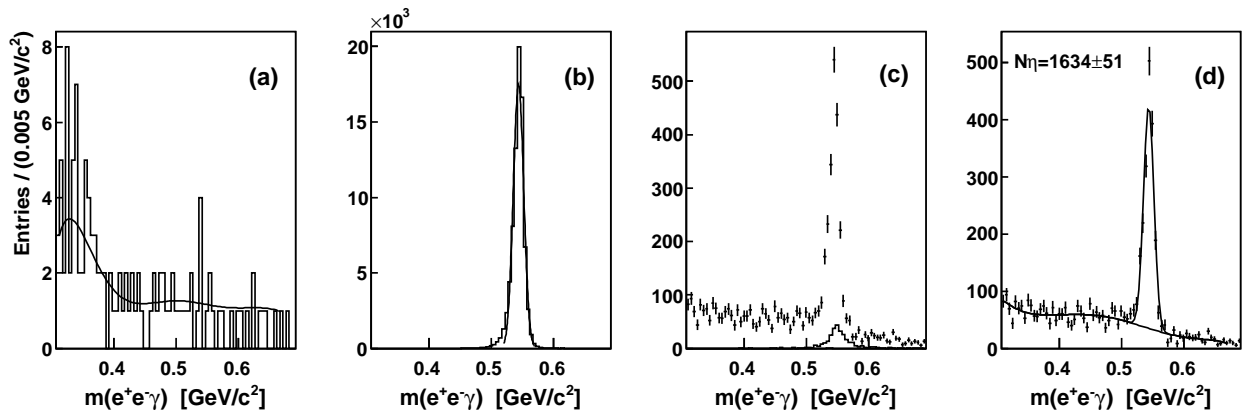}
\caption{
 $m(e^+e^-\gamma)$ invariant-mass distributions obtained for 
 $m(e^+e^-)=(45\pm 5)$~MeV/$c^2$ by using
 both the three- and the four-cluster events:
 (a)~MC simulation of the background reaction $\gamma p\to \pi^0\pi^0p$
     with a polynomial fit;
 (b)~MC simulation of $\gamma p\to \eta p \to e^+e^-\gamma p$
     with a Gaussian fit;
 (c)~experimental events (crosses) after subtracting the random and empty-target
     background; background expected from $\gamma p\to \eta p \to \gamma\gamma p$
     (solid line);
 (d)~experimental events from (c) after subtracting
     the $\gamma p\to \eta p \to \gamma\gamma p$ background
     fitted with the sum of a Gaussian and a polynomial.
}
 \label{fig:eeg34_m045} 
\end{figure*}

 The PID detector was used to separate both the final-state electron and positron
 (the detection efficiency for $e^{+(-)}$ in the PID is close to 100\%)
 from the outgoing photon.
 Since, with respect to the target, the PID provides a full coverage only for
 the CB crystals, events with only three e/m showers in the CB were selected
 for further analysis. Besides improving significantly
 the separation of electrons and positrons from photons with the PID,
 this criterion makes almost all selected events pass the trigger
 requirements (the total energy and the multiplicity in the CB).
 The identification of electrons and positrons was based on a correlation
 between the $\phi$ angles of fired PID elements with the angles
 of e/m showers in the CB.
 The MC simulation of $\gamma p\to \eta p \to e^+e^-\gamma p$ was used
 to optimize this procedure, minimizing a probability of misidentification
 of the photon with either the electron or the positron.
 Such misidentification can occur, for example, if the $\phi$ angle of
 the photon is close to the angle of the electron or the positron. 
 The efficiency of the identification procedure was tested by comparing
 the generated invariant mass $m(e^+e^-)$ with the reconstructed invariant mass.
 In Fig.~\ref{fig:eeg_fig1}(a), the difference between the generated and reconstructed
 $m(e^+e^-)$ values are shown for both three- (dashed line) and four-cluster
 (solid line) events. As seen, the $m(e^+e^-)$ invariant-mass resolution,
 determined by the kinematic-fit technique, is slightly better for
 the four-cluster events, $\sigma_{m}=9.7$~MeV/$c^2$, compared 
 to $\sigma_{m}=10.3$~MeV/$c^2$ for the three-cluster events.
 This is caused by using the outgoing-proton information in the kinematic fit.
 Misidentification of the outgoing photon with either the electron
 or the positron is very small. As found from the MC simulation
 of $\gamma p\to \eta p \to e^+e^-\gamma p$, this misidentification typically occurs
 for high $m(\gamma e^{+(-)})$ masses ($>0.46$~GeV$/c^2$), which correspond
 to highly populated low $m(e^+e^-)$ masses.
 However, the analysis of the experimental data showed that this
 misidentification is not crucial for measuring the $\eta p \to e^+e^-\gamma p$
 signal, as it cannot be determined at $m(e^+e^-)>0.46$~GeV$/c^2$ because of
 a too low signal-to-background ratio.

 The analysis of the MC simulations for all background reactions revealed
 that only the process $\gamma p\to \eta p \to \gamma\gamma p$ could mimic
 the $\eta \to e^+e^-\gamma$ peak. This can occur mostly when 
 one of the final-state photons converts into an electron-positron pair
 in the material between the production vertex and the NaI(Tl) surface,
 or when the photon shower inside the CB splits into two energetic subshowers,
 reconstructed then as two separate clusters (so-called cluster split-offs).
 This background was partially suppressed by optimizing the cluster algorithm 
 and by requiring the number of the fired PID elements to be greater than one.
 The loss of good $\eta \to e^+e^-\gamma$ events because of the PID cut
 is small, as it corresponds to a case when the electron and positron have very close
 $\phi$ angles.  

 Besides the so-called physical background, there are two more background sources.
 The first one comes from interactions of incident photons in the windows
 of the target cell. The subtraction of this background from experimental spectra
 was based on an analysis of data samples that were taken with an empty
 (no liquid hydrogen) target. Another background is due to random coincidences
 of the tagger hits with the experimental trigger; its subtraction was done by using 
 only those tagger hits for which all coincidences were random
 (see Refs.~\cite{slopemamic,etamamic} for more details).
\begin{figure*}
\includegraphics[width=15.5cm,height=5.5cm,bbllx=1.cm,bblly=.25cm,bburx=19.5cm,bbury=6.75cm]{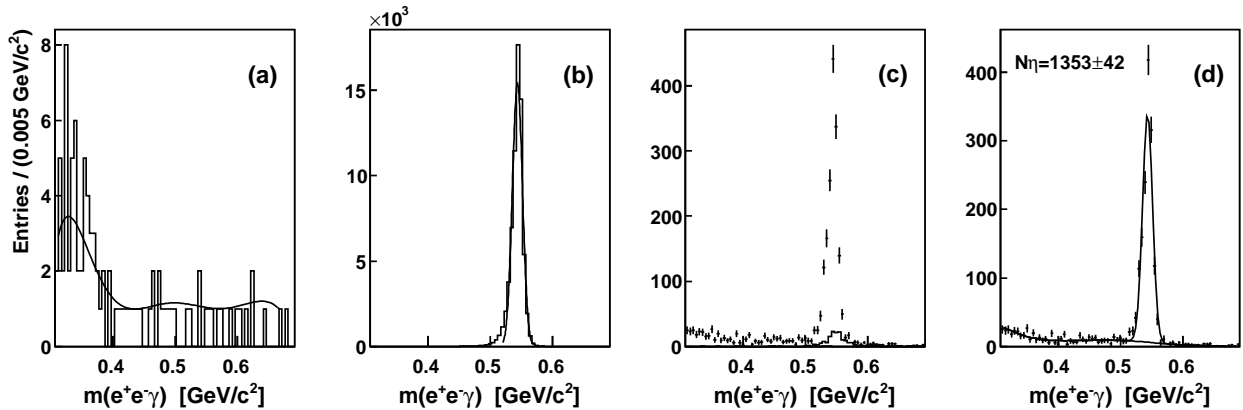}
\caption{
 Same as Fig.~\protect\ref{fig:eeg34_m045}, but
 for the four-cluster events only.
}
 \label{fig:eeg4_m045} 
\end{figure*}
\begin{figure*}
\includegraphics[width=15.5cm,height=5.5cm,bbllx=1.cm,bblly=0.25cm,bburx=19.5cm,bbury=6.75cm]{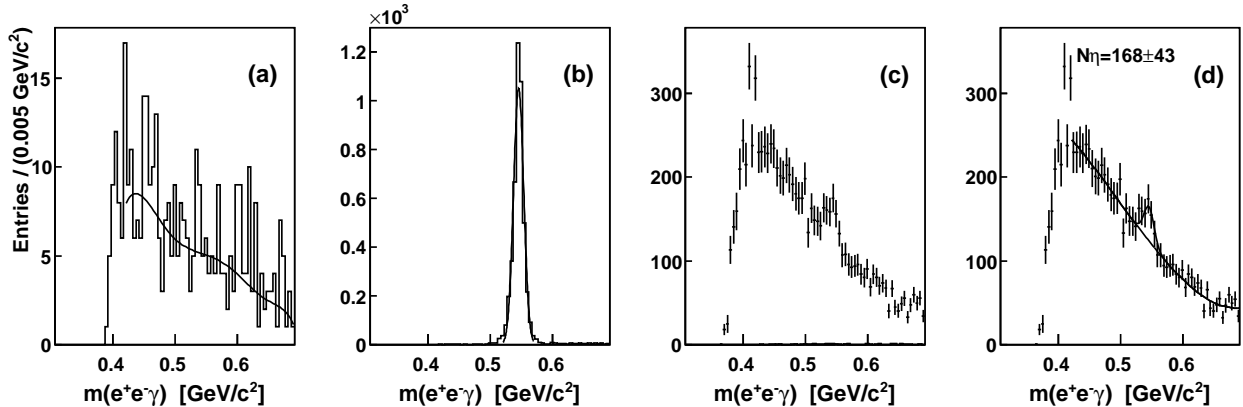}
\caption{
 Same as Fig.~\protect\ref{fig:eeg34_m045}, but
 for $m(e^+e^-)=(370\pm 10)$~MeV/$c^2$.
}
 \label{fig:eeg34_m370} 
\end{figure*}

 In Fig.~\ref{fig:eeg_fig1}(b), the resulting $\gamma p\to \eta p \to e^+e^-\gamma p$
 acceptance is shown as a function of the invariant mass $m(e^+e^-)$ for two cases:
 sum of three- and four-cluster events (solid squares)
 and four-cluster events only (open circles).  
 The $m(e^+e^-)$ acceptance reached for the four-cluster events in the present work
 is about five times better than the acceptance reported in Ref.~\cite{CB_2011}.
 Including the three-cluster events in the analysis gains an additional 20\%
 in the acceptance.
 The acceptance shown in Fig.~\ref{fig:eeg_fig1}(b) is calculated as a ratio
 of the reconstructed events that passed all selection criteria to the all events
 generated for $\gamma p\to \eta p \to e^+e^-\gamma p$.
 Since the $\eta \to e^+e^-\gamma$ decay was generated according to Eq.~(\ref{eqn:dgdm}),
 the number of generated events with large $m(e^+e^-)$ masses is much smaller than
 those with low $m(e^+e^-)$. 
 The increase seen in the $m(e^+e^-)$ acceptance above 0.4~GeV/$c^2$ is artificial;
 it occurs owing to a self-background from the events that have large $m(\gamma e^{+(-)})$
 masses and the outgoing photon is misidentified with the electron or the positron.
 As for the generated $\eta \to e^+e^-\gamma$ decays, large $m(\gamma e^{+(-)})$ masses
 correspond to the highly populated region of low $m(e^+e^-)$ masses, 
 the amount of the self-background events reconstructed with large $m(e^+e^-)$ masses
 is comparable with the number of the actual events at this $m(e^+e^-)$ range.  
 Those self-background events are typically spread wider in the $m(e^+e^-\gamma)$
 distribution and can be partially eliminated by reevaluating the acceptance with
 a Gaussian fit to the peak from $\eta \to e^+e^-\gamma$ decays (how it was done
 in the fitting procedure described later).

 The level of background, remaining under the $\eta \to e^+e^-\gamma$
 signal in the experimental $\gamma p\to e^+e^-\gamma p$ candidates for
 different invariant masses $m(e^+e^-)$,
 can be seen in Figs.~\ref{fig:eeg_fig1}(c) and \ref{fig:eeg_fig1}(d), in which
 the invariant mass $m(e^+e^-)$ is shown
 as a function of the invariant mass $m(e^+e^-\gamma)$
 for all and only the four-cluster events, respectively.
 As seen, despite a lower acceptance,
 the four-cluster events have much smaller background under
 the $\eta \to e^+e^-\gamma$ signal. This becomes especially crucial for
 observing the signal at higher invariant masses $m(e^+e^-)$.
 Another relatively large background is accumulated in the region
 of $m(e^+e^-)$ masses close to the $\pi^0$ mass, which is seen
 in Figs.~\ref{fig:eeg_fig1}(c) and \ref{fig:eeg_fig1}(d) as a horizontal band.
 According to the analysis of the MC simulations for background reactions,
 this band is caused by the $\gamma p\to \pi^0\pi^0p$ background.

 To measure the $\eta \to e^+e^-\gamma$ yield as a function
 of the invariant mass $m(e^+e^-)$, the data were
 divided into several bins along $m(e^+e^-)$.
 Then the peak from the $\eta \to e^+e^-\gamma$ decays was fitted individually
 in every $m(e^+e^-)$ bin for two cases: using all
 $\gamma p\to e^+e^-\gamma p$ candidates and the four-cluster events
 only. The width of the $m(e^+e^-)$ bins was 10 MeV/$c^2$ for
 $m(e^+e^-)<100$~MeV/$c^2$, where the acceptance drops rapidly,
 and 20 MeV/$c^2$ for higher masses.
 The events with $m(e^+e^-)<40$~MeV/$c^2$ were not used
 in the analysis as the acceptance drops very fast in this range.

 In Fig.~\ref{fig:eeg34_m045}, the fitting procedure is illustrated
 for the first bin, $m(e^+e^-)=(45\pm 5)$~MeV/$c^2$, and the case
 when both the three- and four-cluster events are used.
 Figure~\ref{fig:eeg34_m045}(a) depicts the $m(e^+e^-\gamma)$
 invariant-mass distribution for the MC simulation of
 the background reaction $\gamma p\to \pi^0\pi^0p$ fitted with a polynomial.
 Figure~\ref{fig:eeg34_m045}(b) shows a similar distribution
 for the MC simulation of $\gamma p\to \eta p \to e^+e^-\gamma p$
 fitted with a Gaussian.
 The experimental distribution after subtracting the random and empty-target
 background is shown by crosses in Fig.~\ref{fig:eeg34_m045}(c).
 The background remaining from $\gamma p\to \eta p \to \gamma\gamma p$ 
 is shown in the same figure by a solid line. Its normalization
 is based on the number of events generated for
 $\gamma p\to \eta p \to \gamma\gamma p$ and the number
 of the $\gamma p\to \eta p$ events produced in this experiment.
 The experimental distribution after subtraction of
 the $\gamma p\to \eta p \to \gamma\gamma p$ background is
 shown in Fig.~\ref{fig:eeg34_m045}(d).
 A fit to this distribution was done by using the sum of a Gaussian
 and a polynomial. The mean value and $\sigma$ of the Gaussian
 were fixed to the values obtained from the previous fit
 to the MC simulation for $\gamma p\to \eta p \to e^+e^-\gamma p$.
 The initial parameters for the polynomial were taken from
 the fit to the MC simulation for $\gamma p\to \pi^0\pi^0p$.

 To provide a good background description in a quite broad
 range $(0.3-0.7)$~MeV/$c^2$ of the invariant masses $m(e^+e^-\gamma)$
 for all $m(e^+e^-)$ bins and selection criteria,
 an eighth-order polynomial was used for the fits.
 The order was lowered to six for high $m(e^+e^-\gamma)$ masses, where 
 the background range under the $\eta \to e^+e^-\gamma$ signal became narrower.
 It was checked that slight changes of the polynomial order from the used one  
 did not affect the fit results for the $\eta \to e^+e^-\gamma$ signal. 
\begin{figure*}
\includegraphics[width=15.5cm,height=5.5cm,bbllx=1.cm,bblly=.25cm,bburx=19.5cm,bbury=6.75cm]{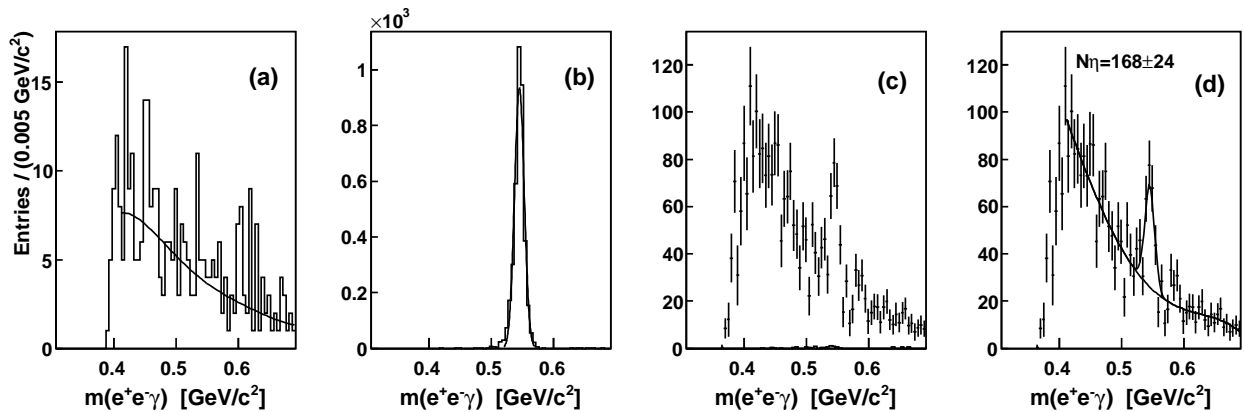}
\caption{
 Same as Fig.~\protect\ref{fig:eeg34_m370}, but
 for the four-cluster events only.
}
 \label{fig:eeg4_m370} 
\end{figure*}

 The experimental number of $\eta \to e^+e^-\gamma$ decays in
 the $m(e^+e^-\gamma)$ distribution shown
 in Fig.~\ref{fig:eeg34_m045}(d) was determined from the area under the Gaussian.
 This number, $1634\pm51$ decays, is already greater than the total number of
 the $\eta \to e^+e^-\gamma$ decays, $1345\pm59$, reported in Ref.~\cite{CB_2011}. 
 Since the calculation of the experimental number of $\eta \to e^+e^-\gamma$ 
 decays was based on the area under a Gaussian,  
 the corresponding detection efficiency was calculated
 in the same way (i.e., based on a Gaussian fit to the MC simulation
 for $\gamma p\to \eta p \to e^+e^-\gamma p$, instead of just using
 the number of entries in the $m(e^+e^-\gamma)$ distribution).

 The fitting procedure for the same $m(e^+e^-)$ bin but using
 only the four-cluster events is illustrated in Fig.~\ref{fig:eeg4_m045}.
 Despite a slightly lower peak from the $\eta \to e^+e^-\gamma$ signal,
 a significant improvement of the signal-to-background ratio can be seen.

 The fits to both the three- and the four-cluster events
 were made only up to $m(e^+e^-)=400$~MeV/$c^2$.
 Above this energy, those fits became unreliable as the signal from $\eta \to e^+e^-\gamma$
 decays became comparable with statistical fluctuations of the background. 
 For the four-cluster events, fitting the peak from $\eta \to e^+e^-\gamma$ decays
 was possible up to $m(e^+e^-)=460$~MeV/$c^2$.
 In Fig.~\ref{fig:eeg34_m370}, the fitting procedure is illustrated
 for the range $m(e^+e^-)=(370\pm 10)$~MeV/$c^2$, in which a peak from
 $\eta \to e^+e^-\gamma$ decays is still clearly seen for the case
 when both the three- and the four-cluster events are used.
 The corresponding fit to only the four-cluster events is shown
 in Fig.~\ref{fig:eeg4_m370}.
 As seen in Figs.~\ref{fig:eeg34_m370} and \ref{fig:eeg4_m370},
 the $\gamma p\to \eta p \to \gamma\gamma p$ background is
 negligibly small in this range of $m(e^+e^-)$, whereas the
 background under the peak from $\eta \to e^+e^-\gamma$ decays increases
 greatly. Also, the fit yields a sufficiently larger
 uncertainty in the number of the $\eta \to e^+e^-\gamma$ decays
 for the case of a larger background, i.e., when both the three-
 and four-cluster events are used. 

\section{Discussion of the results}
  \label{sec:Results}

\begin{figure*}
\includegraphics[width=15.5cm,height=7.cm,bbllx=1.cm,bblly=.0cm,bburx=19.5cm,bbury=7.8cm]{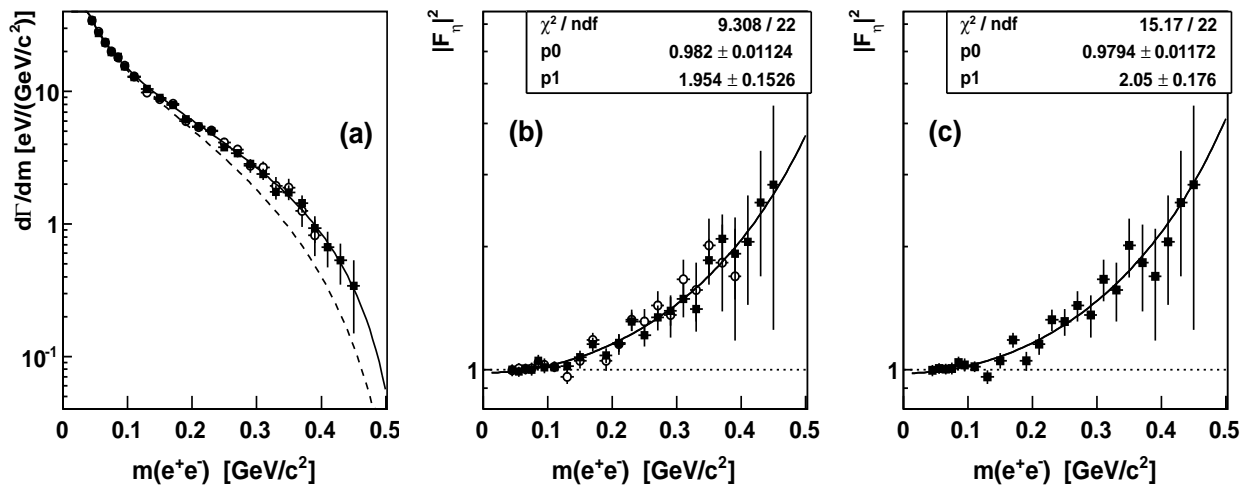}
\caption{
 (a)~Results for $d\Gamma(\eta \to e^+e^-\gamma)/dm(e^+e^-)$ obtained
 by using the four-cluster events only (solid squares) and
 both the three- and the four-cluster events (open circles).
 The QED prediction with $|F_{\eta}|^2=1$ and the QED prediction including
 the $\eta$ TFF with $\Lambda^{-2}=1.95$~GeV$^{-2}$ are shown by a dashed
 and a solid line, respectively.
 ~(b)~ Main results (based on the four-cluster events only) of this work
   (solid squares) for $|F_{\eta}(m_{e^+e^-})|^2$ compared to
   the results based on all events (open circles); the solid line shows
   a fit of the $\eta$ TFF parametrized as Eq.~(\protect\ref{eqn:Fm})
   to the main results.
 ~(c)~ Fit (solid line) to results for $|F_{\eta}(m_{e^+e^-})|^2$
       obtained from combining the three- and four-cluster events, and
      used then to estimate the systematic uncertainty in the $\Lambda^{-2}$ value.
}
 \label{fig:dgdm_etaeeg_3x1} 
\end{figure*}

 The number of the $\eta \to e^+e^-\gamma$ decays initially produced
 in each $m(e^+e^-)$ bin was obtained by
 dividing the value from a Gaussian fit to the experimental distribution
 by the corresponding detection efficiency (see Sec.~\ref{sec:Data}
 for details). 
 Values for $d\Gamma(\eta \to e^+e^-\gamma)/dm(e^+e^-)$ were obtained
 by using the full decay width $\Gamma_{\eta}=1.30$~keV~\cite{PDG} and the
 total number of $\eta$ mesons produced, which was determined
 from an analysis of the process $\gamma p\to \eta p \to 3\pi^0p$
 in the same data set~\cite{etamamic} and using 0.3257 for
 the $\eta \to 3\pi^0$ branching ratio~\cite{PDG}. 
 The results for $d\Gamma(\eta \to e^+e^-\gamma)/dm(e^+e^-)$
 obtained by using only the four-cluster events are shown
 by solid squares in Fig.~\ref{fig:dgdm_etaeeg_3x1}(a).
 The corresponding results obtained for the sum of the three- and
 four-cluster events are shown in the same figure by open circles. 
\begin{table*}
\caption
[tab:etatff]{
 Results of this experiment for the $\eta$ TFF, $|F_{\eta}|^2$, as a function of
 the invariant mass $m(e^+e^-)$
 } \label{tab:etatff}
\begin{ruledtabular}
\begin{tabular}{|c|c|c|c|c|c|c|} 
\hline
 $m(e^+e^-)$~[MeV/$c^2$]
 & $45\pm5$ & $55\pm5$ & $65\pm5$ & $75\pm5$ & $85\pm5$ & $95\pm5$ \\
\hline
 $|F_{\eta}|^2$ 
 & $0.999\pm0.031$ & $0.988\pm0.029$ & $1.005\pm0.030$ & $0.999\pm0.031$
 & $1.051\pm0.034$ & $1.014\pm0.036$ \\
\hline
\hline
 $m(e^+e^-)$~[MeV/$c^2$]
 & $110\pm10$ & $130\pm10$ & $150\pm10$ & $170\pm10$ & $190\pm10$ & $210\pm10$ \\
\hline
 $|F_{\eta}|^2$ 
 & $1.014\pm0.028$ & $1.019\pm0.037$ & $1.071\pm0.041$ & $1.153\pm0.044$
 & $1.083\pm0.046$ & $1.161\pm0.056$ \\
\hline
\hline
 $m(e^+e^-)$~[MeV/$c^2$]
  & $230\pm10$ & $250\pm10$ & $270\pm10$ & $290\pm10$  & $310\pm10$ & $330\pm10$ \\
\hline
 $|F_{\eta}|^2$ 
 & $1.312\pm0.068$ & $1.214\pm0.076$ & $1.342\pm0.094$ & $1.393\pm0.113$
 & $1.487\pm0.144$ & $1.406\pm0.170$ \\
\hline
\hline
 $m(e^+e^-)$~[MeV/$c^2$]
 & $350\pm10$ & $370\pm10$ & $390\pm10$ & $410\pm10$  & $430\pm10$ & $450\pm10$ \\
\hline
 $|F_{\eta}|^2$ 
 & $1.851\pm0.235$ & $2.086\pm0.306$ & $1.918\pm0.433$ & $2.05\pm0.61$
 & $2.56\pm0.87$ & $2.83\pm1.58$ \\
\hline
\end{tabular}
\end{ruledtabular}
\end{table*}
 As seen, in the range of the latter results, $m(e^+e^-)<400$~MeV/$c^2$,
 the $d\Gamma(\eta \to e^+e^-\gamma)/dm(e^+e^-)$ values
 of both the approaches are in good agreement within their error bars.
 The QED prediction for $d\Gamma(\eta \to l^+l^-\gamma)/dm(l^+l^-)$ with
 $|F_{\eta}|^2=1$ is depicted in Fig.~\ref{fig:dgdm_etaeeg_3x1}(a) by
 a dashed line, and the QED prediction including the $\eta$ TFF with
 $\Lambda^{-2}=1.95$~GeV$^{-2}$ (most precise experimental result
 from the NA60~\cite{NA60_2009,NA60_2012}) is shown by a solid line.
 One can see that the results of this work are much closer to
 the prediction based on the NA60 data.   
\begin{figure*}
\includegraphics[width=15.cm,height=8.cm,bbllx=1.5cm,bblly=0.2cm,bburx=19.cm,bbury=8.5cm]{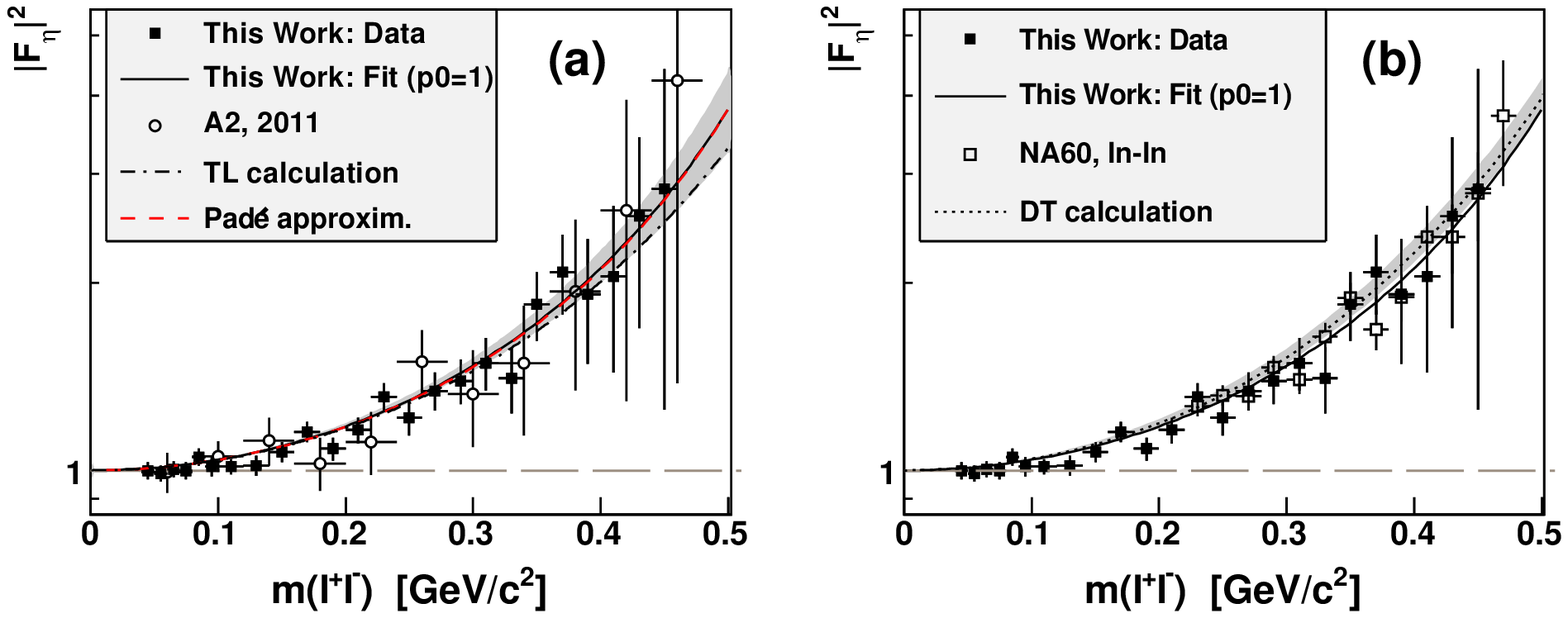}
\caption{ (Color online)
 Results of this work (solid squares) for the $\eta$ TFF, $|F_{\eta}(m_{ll})|^2$,
 compared to other recent measurements and theoretical predictions:
 former data of the A2 Collaboration~\protect\cite{CB_2011} (open circles in (a))
 and the NA60 in peripheral In--In data~\protect\cite{NA60_2009} (open squares in (b)),
 calculations of Ref.~\protect\cite{Ter10} (dash-dotted line in (a)),
 Ref.~\protect\cite{Esc13} (red dashed line with an error band in (a)),
 and Ref.~\protect\cite{Hanhart} (dotted line with an error band in (b)).
 The solid line is the fit from Fig.~\protect\ref{fig:dgdm_etaeeg_3x1}(b)
 rescaled so that $p0=1$.
}
 \label{fig:eta_tff_exp_thr} 
\end{figure*}

 The results for the $\eta$ TFF, $|F_{\eta}|^2$, were obtained
 by dividing the $d\Gamma(\eta \to e^+e^-\gamma)/dm(e^+e^-)$ values
 by the QED term of Eq.~(\ref{eqn:dgdm}) and using 0.3931 for
 the $\eta \to \gamma\gamma$ branching ratio~\cite{PDG}.
 In Fig.~\ref{fig:dgdm_etaeeg_3x1}(b), the $|F_{\eta}(m_{ll})|^2$ values
 are shown by solid squares for the case of using 
 only the four-cluster events and by open circles for all events. 
 As seen, the TFF results of both the approaches
 are in good agreement within their error bars.
 Because of the background under the $\eta \to e^+e^-\gamma$ signal,
 the uncertainties of the data points do not reflect
 the actual statistic for the $\eta \to e^+e^-\gamma$ decays
 observed, which was $1.8\cdot 10^4$ and $2.2\cdot 10^4$ 
 for only four-cluster and all events, respectively.
 Some staggering of the data points is also caused by
 statistical fluctuations of background events under
 the $\eta \to e^+e^-\gamma$ peak.

 Since the data based only on the four-cluster events
 have a better signal-to-background ratio, resulting in smaller fit uncertainties,
 and a wider covering of the $m(e^+e^-)$ range, they
 are considered as the main results of this work.
 Their numerical values are listed in Table~\ref{tab:etatff}.
 The magnitude of $\Lambda^{-2}$ was obtained by fitting
 the $\eta$ TFF parametrized as defined in Eq.~(\ref{eqn:Fm})
 to the main $|F_{\eta}|^2$ results.
 As discussed in the Introduction, the fit was made
 with two free parameters, one of which, $p1$,
 is $\Lambda^{-2}$ itself, and the other, $p0$, reflects
 the general normalization of the data points,
 which could be different from $p0=1$ because of
 the uncertainty in the determination of
 the experimental number of $\eta$ mesons produced.
 The correlation between the two parameters results
 in a larger fit error for $\Lambda^{-2}$.
 However, this fit error already includes the systematic uncertainty
 in the general normalization of the data points.

 The two-parameter fit to the main $|F_{\eta}|^2$ data points is
 shown in Fig.~\ref{fig:dgdm_etaeeg_3x1}(b) by a solid line.
 This fit yields the values $p0=0.982\pm0.011$ for the normalization parameter
 and $p1=(1.95\pm0.15)$~GeV$^{-2}$ for $\Lambda^{-2}$. For simplicity,
 the fit error for $\Lambda^{-2}$ will be called its statistical uncertainty
 throughout the rest of this paper.

 The results based on the sum of the three- and four-cluster events
 were used to estimate the systematic uncertainty that
 comes from fitting the background under the $\eta \to e^+e^-\gamma$ peak
 (as the two subsets have different levels of background) and
 from the acceptance correction, including the detection efficiency
 for the outgoing proton.
 This systematic uncertainty was estimated by replacing 
 the main $|F_{\eta}|^2$ results below $m(e^+e^-)=400$~MeV/$c^2$
 with the results obtained from the sum of the three- and four-cluster events and
 repeating the fit with the $\eta$ TFF parametrized as Eq.~(\ref{eqn:Fm}).
 This fit, demonstrated in Fig.~\ref{fig:dgdm_etaeeg_3x1}(c),
 yields $\Lambda^{-2}=(2.05\pm 0.18)$~GeV$^{-2}$, which, within the uncertainties, is
 in good agreement with the value obtained from the fit to the main results.
 The difference between the two results was taken as the systematic
 uncertainty in the value of $\Lambda^{-2}$ measured in this work.
 Then, the final value for the present measurement is
\begin{equation}
 \Lambda^{-2}=(1.95\pm 0.15_{\mathrm{stat}}\pm 0.10_{\mathrm{syst}}) ~\mathrm{GeV}^{-2},
\label{eqn:Lam2_this_work}
\end{equation}
 which is in very good agreement within the errors with all
 recent results reported in Refs.~\cite{CB_2011,NA60_2009,NA60_2012}.
 As seen in Fig.~\ref{fig:eta_tff_exp_thr}, the $|F_{\eta}(m_{ll})|^2$ results
 of this work are in similar good agreement within the error bars
 with the data points from Refs.~\cite{CB_2011,NA60_2009}.
 
 The uncertainty reached for the $\Lambda^{-2}$ value
 in the present work is smaller than those of 
 all previous measurements based on the $\eta \to e^+e^-\gamma$ decay,
 is of a similar magnitude as the NA60 value from peripheral In--In data~\cite{NA60_2009},
 and still yields to the latest, preliminary result of the NA60
 from $p-A$ collisions~\cite{NA60_2012}.

 In Fig.~\ref{fig:eta_tff_exp_thr}, the results of this work
 for $|F_{\eta}(m_{ll})|^2$ are also compared to three different theoretical predictions.
 Since all models assume that $|F_{\eta}(m_{ll}=0)|^2=1$,
 for a better comparison, the fit to the data points from
 Fig.~\ref{fig:dgdm_etaeeg_3x1}(b) is rescaled by setting
 its normalization parameter to $p0=1$ and leaving its second parameter $p1$,
 reflecting the slope parameter $\Lambda^{-2}$, unchanged.
 The calculation by Terschl\"usen and Leupold (TL) combines the vector-meson
 Lagrangian proposed in Ref.~\cite{Lut08} and recently extended in Ref.~\cite{Ter12},
 with the Wess-Zumino-Witten contact interaction~\cite{Ter10}
 (see also Ref.~\cite{Ter13} for the corresponding case of the $\pi^0$ TFF).
 Their calculation agrees very well with the standard VMD form factor.
 As seen, the TL calculation (shown in Fig.~\ref{fig:eta_tff_exp_thr}(a)
 by a dash-dotted line) goes slightly lower than the pole-approximation
 (Eq.~(\ref{eqn:Fm})) fit to the present data, whereas it fully describes
 the data points within the error bars.

 The second calculation is based on a model-independent method
 using Pad\'e approximants that was developed for the $\pi^0$ TFF in Ref.~\cite{Mas12}.
 Using spacelike data (CELLO~\cite{Behrend}, CLEO~\cite{Gronberg},
 BABAR~\cite{Amo_Sanchez}), this method provides a parametrization
 that is also suited to describe data in the $m_{ll}$ range
 from zero to $\sqrt{0.4})$~GeV/$c^2$,
 and thus provides a model-independent prediction for the timelike TFF~\cite{Esc13}.
 Over the full $m_{ll}$ range, this calculation (shown
 in Fig.~\ref{fig:eta_tff_exp_thr}(a) by a red dashed line with
 an error band) practically overlaps with the pole-approximation fit to the present
 data points.

 In another recent calculation~\cite{Hanhart} by the J\"ulich group,
 the connection between the radiative decay
 $\eta \to \pi^+ \pi^- \gamma$ and the isovector contributions of
 the $\eta \to \gamma \gamma^*$ TFF is exploited in a model-independent way,
 using dispersion theory (DT).  This calculation (shown in
 Fig.~\ref{fig:eta_tff_exp_thr}(b) by a dotted line with an error band) goes
 slightly above the fit to the present data.

 Currently, the VMD models that are used to calculate
 the contribution of the hadronic light-by-light scattering
 to $(g-2)_\mu$ include only $\rho$, $\omega$, and $\phi$ resonances.
 These contributions are calculated with $\Lambda=(774 \pm 29)$~MeV close to
 the $\rho$-meson mass. This value of $\Lambda$ was determined from a fit
 to spacelike data measured by the CLEO collaboration~\cite{Gronberg}
 down to the momentum transfer $q^2 = -1.5$~GeV$^2$, which is
 far away from $q^2=0$~GeV$^2$.
 The $\Lambda$ value from CLEO disagrees with the VMD value, $\Lambda=745$~MeV.
 It also disagrees with the result of this work, $\Lambda=(716 \pm 0.033)$~MeV,
 by more than one standard deviation, where both the measurements are
 of a similar accuracy. However, the measurement presented in this work
 was performed much closer to $q^2=0$~GeV$^2$, and it agrees very well
 with earlier measurements of the $\eta\to e^+ e^- \gamma$ Dalitz decay,
 just improving their accuracy.
 Though the results of this work are not able to rule out the VMD models used
 for calculating $(g-2)_\mu$, one can see that smaller values for $\Lambda$
 should be used in those calculations, indicating that contributions
 from heavier vector-meson resonances, like, e.g., $\rho'$,
 might not be completely negligible.
 
\section{Summary and conclusions}
\label{sec:Conclusion}

 A new determination of the electromagnetic transition form factor from
 the $\eta \to e^+ e^- \gamma$ Dalitz decay was presented in this paper.
 The statistical accuracy achieved in this work surpasses all previous measurements
 of $\eta \to e^+ e^- \gamma$ and matches the NA60 result based on
 $\eta \to \mu^+ \mu^- \gamma$ decays from peripheral In-In collisions.
 Compared to the former determination of the $\eta$ TFF by
 the A2 Collaboration, an increase
 by more than one order of magnitude in statistic has been achieved.
 This was accomplished by an analysis of three times more data and the use
 of a kinematic-fit technique, which allowed for far looser cuts and
 for exploiting the full $\eta$ production range available at MAMI-C.
 The extracted slope parameter
 $\Lambda^{-2} = (1.95\pm 0.15_{\mathrm{stat}}\pm 0.10_{\mathrm{syst}})$~GeV$^{-2}$
 agrees within the uncertainties with the results from all recent measurements
 of the $\eta$ TFF. A pole-approximation fit to the presented data shows
 almost perfect agreement with the model-independent calculation
 from Ref.~\cite{Esc13}.
 The calculation by Terschl\"usen and Leupold~\cite{Ter10}
 and the DT calculation~\cite{Hanhart} deviate slightly from the fit,
 but the statistical uncertainties are still not sufficient to rule out
 any of the theoretical predictions.
 Thus, a need for more precise measurements is evident,
 though the results of this work indicate clearly
 that smaller effective vector-meson masses $\Lambda$ should be used in
 VMD-like models for calculating the contribution of the hadronic light-by-light
 scattering to $(g-2)_\mu$, as well as rare $\eta$ decays and
 all processes involving the $\eta$ TFF.

\section*{Acknowledgment}

 The authors wish to acknowledge the excellent support of the accelerator group and
 operators of MAMI. We would like to thank
 Pablo Sanchez-Puertas for many fruitful discussions.
 This work was supported by the Deutsche Forschungsgemeinschaft (SFB443,
 SFB/TR16, and SFB1044), DFG-RFBR (Grant No. 09-02-91330), the European Community-Research
 Infrastructure Activity under the FP6 ``Structuring the European Research Area''
 program (Hadron Physics, Contract No. RII3-CT-2004-506078), Schweizerischer
 Nationalfonds, the U.K. Science and Technology Facilities Council, the U.S. Department
 of Energy and National Science Foundation, INFN (Italy), and NSERC (Canada).
 The work of P.~Masjuan was supported by the Cluster of excellence ``PRISMA''
 of the Deutsche Forschungsgemeinschaft and the State of Rhineland Palatinate, Germany.
 A.~Fix acknowledges additional support from the Russian Federation federal program
 ``Kadry'' (Contract No. P691) and the MSE Program ``Nauka'' (Contract No. 1.604.2011).
 We thank the undergraduate students of Mount Allison University
 and The George Washington University for their assistance.

\end{document}